\documentclass[usenatbib]{emulateapj}
\usepackage{graphicx}
\usepackage[flushleft]{threeparttable}
\usepackage[usenames,dvipsnames,svgnames,table]{xcolor}
\usepackage{hyperref}
\definecolor{darkblue}{rgb}{0.0,0.0,0.3}
\hypersetup{colorlinks,breaklinks,
            linkcolor=darkblue,urlcolor=darkblue,
            anchorcolor=darkblue,citecolor=darkblue}
\usepackage{amsmath,amssymb}
\usepackage{amsmath}
\usepackage{amsfonts}

\usepackage{natbib}
\makeatletter
\renewcommand{\p@subsection}{}
\renewcommand{\p@subsubsection}{}
\makeatother

\begin{document}

\def\etal{et al.\ \rm}
\def\ba{\begin{eqnarray}}
\def\ea{\end{eqnarray}}
\def\etal{et al.\ \rm}
\def\Fdw{F_{\rm dw}}
\def\Tex{T_{\rm ex}}
\def\Fdis{F_{\rm dw,dis}}
\def\Fnu{F_\nu}
\def\WD{\rm WD}

\newcommand\cmtrr[1]{{\color{red}[RR: #1]}}

\newcommand\rrr[1]{{\bf\color{red}#1}}

%%%%%%%%%%%%%%%%%%%%%%%%%%%%%%%%%%%%%%%%%%%%%%%%%%%%%%%%%%%

\title{Doppler Boosting of the S-stars in the Galactic Center}

\author{Roman R. Rafikov\altaffilmark{1,2,3}}
\altaffiltext{1}{Institute for Advanced Study, Einstein Drive, Princeton, NJ 08540}
\altaffiltext{2}{Centre for Mathematical Sciences, Department of Applied Mathematics and Theoretical Physics, University of Cambridge, Wilberforce Road, Cambridge CB3 0WA, UK; rrr@damtp.cam.ac.uk}
\altaffiltext{3}{John N. Bahcall Fellow at the Institute for Advanced Study}

%%%%%%%%%%%%%%%%%%%%%%%%%%%%%%%%%%%%%%%%%%%%%%%%%%%%%%%%%%%

\begin{abstract}
Astrometry and spectroscopy of the S-stars in the Galactic Center provide a unique way to probe the properties of the central supermassive black hole, as well as the post-Newtonian effects caused by its gravity, e.g. gravitational redshift and general relativistic precession. It has also been suggested that the photometry of S-stars can be used for studying the properties of gaseous environment of Sgr A$^*$. Due to the high velocities of the S-stars, sometimes approaching $~0.1c$, their photometric signal should be considerably affected by the Doppler boosting. We calculate this relativistic effect for several S-stars closely approaching the central black hole (most of them recently announced) and show that the amplitude of the photometric variability due to the Doppler boosting for some of them (S62 and S4714) exceeds $6\%$; for the well studied star S2 it is about $2\%$. Measurement of the Doppler boosting can confirm the existence and help refine orbital parameters of the S-stars with noisy spectroscopy and astrometry. This effect should be explicitly accounted for when the photometry of S-stars is used for probing the medium around the Sgr A$^*$. We discuss the observability of the Doppler boosting given the complications typical for the Galactic Center and conclude, in particular, that the purely photometric detection of the higher order relativistic corrections to the Doppler boosting signal (due to the gravitational redshift and transverse Doppler shift, which we also calculate) is hardly possible for the S-stars.
\end{abstract}

%%%%%%%%%%%%%%%%%%%%%%%%%%%%%%%%%%%%%%%%%%%%%%%%%%%%%%%%%%%

%%%%%%%%%%%%%%%%%%%%%%%%%%%%%%%%%%%%%%%%%%%%%%%%%%%%%%%%%%%
%%%%%%%%%%%%%%%%%%%%%%%%%%%%%%%%%%%%%%%%%%%%%%%%%%%%%%%%%%%

\section{Introduction.}  
\label{sect:intro}

%%%%%%%%%%%%%%%%%%%%%%%%%%%%%%%%%%%%%%%%%%%%%%%%%%%%%%%%%%%

S-stars are a population of young, bright stars orbiting within 0.04 pc from the supermassive black hole (BH) in the Galactic Center \citep{Habibi2017}. Long-term monitoring of these stars reveal their Keplerian motion around the central BH with periods as short as tens of years and, in many cases, high eccentricities \citep{Ghez2003b,Gil2017}. Observations of one particular member of the S-cluster --- the relatively bright (m$_K\approx 14$) star S2 making a full orbit in about 16 yr and approaching the BH as close as $115$ au \citep{Sch2002,Ghez2003a} --- allowed accurate measurements of the BH mass and the distance to the Galactic Center \citep{Ghez2000,Eisen2003}, as well as the detection of some post-Newtonian effects, namely the gravitational redshift \citep{Gravity2018redshift,Do2019} and general relativistic (GR) precession \citep{Gravity2020prec}. Recently, several fainter S-stars approaching Sgr A$^*$ (a source centered on the BH) even closer than S2 have been announced by \citet{Peis2020a,Peis2020b}.

S-stars have been extensively studied using spectroscopy (to obtain their radial velocities) and astrometry to understand their motion in three dimensions. Recently, \citet{Elaheh2020} proposed to also use the {\it photometry} of S2 as a probe of the gaseous environment of Sgr A$^*$. They suggested that the photometric signal of S2 may be perturbed by the bow shock that the star drives in the gaseous environment of Sgr A$^*$, via either thermal or synchrotron emission at the shock. Although \citet{Elaheh2020} found no evidence for the photometric variability of S2 at the level of $2.5\%$ (setting indirect constraints on the gas density near its pericenter), they have added stellar photometry to the toolbox of methods used to study the S-stars.

Due to the very high velocities of S-stars, in some cases reaching  $\sim 0.1c$ at the pericenter, their brightness should be significantly modified by the phenomenon of Doppler boosting (also known as Doppler or relativistic beaming) --- the process by which the relativistic effects change the apparent brightness of the moving object. This effect has been previously invoked for a variety of astrophysical objects, e.g. relativistic jets \citep{Begel1984}, compact binaries \citep{Shak1987,van2010}, stars hosting short-period exoplanets \citep{Loeb2003,Mazeh2010}, and supermassive black hole binaries with accretion disks \citet{Dorazio2015}. Here, prompted in part by the recently announced new S-stars closely approaching the Galactic Center black hole \citep{Peis2020a,Peis2020b}, we demonstrate the importance of the Doppler boosting for some of these objects\footnote{Potential role of the Doppler boosting in the Galactic Center was briefly mentioned in \citet{Zucker2006}, but then it was dismissed as hardly observable in \citet{Zucker2007}.} and show that the associated photometric signal can reach significant levels (several per cent). 

%%%%%%%%%%%%%%%%%%%%%%%%%%%%%%%%%%%%%%%%%%%%%%%%%%%%%%%%%%%
%%%%%%%%%%%%%%%%%%%%%%%%%%%%%%%%%%%%%%%%%%%%%%%%%%%%%%%%%%%

\section{Doppler boosting near the Galactic Center Black Hole}  
\label{sect:boosting}

%%%%%%%%%%%%%%%%%%%%%%%%%%%%%%%%%%%%%%%%%%%%%%%%%%%%%%%%%%%

As the star orbits the Galactic BH, its flux received by a distant observer changes due to both special and general relativistic effects --- time dilation, aberration of light, change of the photon frequency due to the stellar motion and gravitational redshift. The combination of these effects results in Doppler boosting of the stellar emission.

%%%%%%%%%%%%%%%%%%%%%%%%%%%
\begin{table*}
\begin{threeparttable}
\caption{Assumed orbital parameters and computed variability characteristics of the S-stars used in this work}
\begin{tabular}{llllllllllc|cc}
    \hline \hline\\
    Source & mag$_K$ & $a$ (mpc) & $e$ & $i$ ($^\circ$) & $\omega$ ($^\circ$) & $\Omega$ ($^\circ$) & $T_0$ (yr) & $P_{\rm orb}$ (yr)  & $r_{\rm p}$ (au) & Ref & $\delta(\Delta F/F_{\rm em}), \%$ & $\delta(\Delta F/F_{\rm em})_2, \%$ 
    \\
    \\
    \hline
    \\
S2 & 14.1 & 4.895 & 0.886 & 133.9 & 66.0 & 227.4 & 2018.377 & 16.04 & 115.1 & 1 & 1.96 & 0.061\\ 
    %\hline
S62 & 16.1 & 3.59 & 0.976 & 72.7 & 42.6 & 122.6 & 2003.33 &10.08 & 17.8 & 2\tnote{*} & 6.43 & 0.31\\
    %\hline
S4711 & 18.4 & 3.0 & 0.77 & 114.7 & 131.6 & 20.1 & 2010.85 & 7.7 & 142.4 & 2 & 2.27 & 0.04 \\
    %\hline
S4714 & 17.7 & 4.08 & 0.985 & 127.7 & 357.3 & 129.3 & 2017.29 & 12.21 & 12.6 & 2 & 6.49 & 0.25 \\
\\
    \hline   
\end{tabular}
\begin{tablenotes}
\item {\bf Notes}: For every star we list our adopted central values (see references for error bars) for its $K$-magnitude, semi-major axis $a$, eccentricity $e$, inclination $i$  (relative to the line of sight), pericenter angle $\omega$, nodal angle $\Omega$, time of pericenter passage $T_0$, orbital period $P_{\rm orb}$ (assuming BH mass $M_{\rm BH}=3.98\times 10^6$ M$_\odot$, \citealt{Do2019}), and pericenter distance $r_{\rm p}$. We also list the calculated range of the photometric variability due to Doppler boosting $\delta(\Delta F/F_{\rm em})$ (Eq. (\ref{eq:final})), as well as the range $\delta(\Delta F/F_{\rm em})_2$ of the contribution due to the transverse Doppler shift and gravitational redshift (Eq. (\ref{eq:final2})). References: (1) \citet{Do2019}, (2) \citet{Peis2020b}.
\item[*] \citet{Gravity2020-S62} challenged the determination of the S62 orbital parameters by \citet{Peis2020a,Peis2020b}, which we still use here to provide an example.
\end{tablenotes}
\label{table:stars}
\end{threeparttable}
\end{table*}
%%%%%%%%%%%%%%%%%%%%%%%%%%%

We will be interested in the stellar flux received by the observer at rest far from the Galactic Center (corrections due to the Earth motion and non-zero gravitational potential of the observer can be trivially added, \citealt{Blanchet2001}) in a narrow filter with the window function $w(\nu)$  and central frequency $\nu_{\rm obs}$. If $I_{\nu,{\rm obs}}$ is the radiation intensity in the observer's frame, then the flux in that filter measured by the observer is
\ba
F_{\rm obs}=\int\limits_0^\infty I_{\nu,{\rm obs}}(\nu_{\rm obs}) w(\nu_{\rm obs}){\rm d}\nu_{\rm obs}.
\label{eq:master}
\ea
Photon number conservation dictates that $I_\nu(\nu)/\nu^3$ is a relativistic invariant \citep{Ryb}, meaning that $I_{\nu,{\rm obs}}(\nu_{\rm obs})$ is related to the intensity in the stellar (emitter) frame $I_{\nu,{\rm em}}(\nu_{\rm em})$, at the frequency of emission $\nu_{\rm em}$ related to $\nu_{\rm obs}$ through the Doppler shift, as $I_{\nu,{\rm obs}}(\nu_{\rm obs})=I_{\nu,{\rm em}}(\nu_{\rm em})(\nu_{\rm obs}/\nu_{\rm em})^3$. Assuming that the filter is narrow and that $\nu_{\rm obs}/\nu_{\rm em}$ only slightly deviates from unity, we can also write 
\ba  
I_{\nu,{\rm em}}(\nu_{\rm em})\approx I_{\nu,{\rm em}}(\nu_{\rm obs})\left(\nu_{\rm em}/\nu_{\rm obs}\right)^\alpha,
\label{eq:approx}
\ea   
where
\ba  
\alpha=\left({\rm d}\ln I_{\nu,{\rm em}}/{\rm d}\ln\nu\right)\big|_{\nu=\nu_{\rm obs}}.
\label{eq:alpha}
\ea  
Plugging these relations into equation (\ref{eq:master}) and using the fact that $\nu_{\rm obs}/\nu_{\rm em}$ is independent of $\nu_{\rm obs}$ (see Eq. \ref{eq:nu-rat} below), one obtains 
\ba  
F_{\rm obs}=F_{\rm em}\left(\nu_{\rm obs}/\nu_{\rm em}\right)^{3-\alpha},
\label{eq:flux-rat}
\ea  
where the emitted flux, assumed to not vary in time, is $F_{\rm em}=\int_0^\infty I_{\nu,{\rm em}}(\nu) w(\nu){\rm d}\nu$.

To relate $\nu_{\rm em}$ to $\nu_{\rm obs}$ we use the results of \citet{Blanchet2001}, which account for both special and general relativistic effects and are accurate up to $O(v^2/c^2)$:
\ba  
\frac{\nu_{\rm em}}{\nu_{\rm obs}}\approx
\frac{1+v_{\rm LOS}/c}{1+\left(\Phi_{\rm BH}(r)-v^2/2\right)/c^2},
\label{eq:nu-rat}
\ea  
where $v_{\rm LOS}$ is the stellar line-of-sight velocity (assumed positive for {\it receding} objects) and $v$ is the full velocity of the emitting star, while $\Phi_{\rm BH}(r)=-GM_{\rm BH}/r$ is the BH potential at the instantaneous stellar distance $r$ ($M_{\rm BH}\approx 3.98\times 10^6$ M$_\odot$ is the BH mass, \citealt{Do2019}). In this expression the terms in the denominator multiplied by $c^{-2}$ represent the $O(v^2/c^2)$ correction due to the general relativistic redshift and the special relativistic transverse Doppler shift, respectively. They go beyond the standard $O(v/c)$ non-relativistic Doppler shift expression appearing in the numerator, and we explore if they can provide a measurable contribution to the stellar photometric signal in \S \ref{sect:results}. 

Equations (\ref{eq:flux-rat}) and (\ref{eq:nu-rat}) yield the following expression for the relative flux variation due to the Doppler boosting 
\ba  
\frac{\Delta F}{F_{\rm em}}&=&\frac{F_{\rm obs}-F_{\rm em}}{F_{\rm em}}
\nonumber\\
&=& \left[\frac{1+v_{\rm LOS}/c}{1+\left(\Phi_{\rm BH}(r)-v^2\right/2)/c^2}\right]^{\alpha-3}-1,
\label{eq:final}
\ea  
which is valid to $O(v^2/c^2)$ order. Retaining the terms only to first order in $v/c\ll 1$ one recovers the familiar result \citep{Loeb2003,Dorazio2015}
\ba  
\left(\Delta F/F_{\rm em}\right)_1\approx  (\alpha-3)\left(v_{\rm LOS}/c\right)
\label{eq:final1}
\ea  
for the flux variation $\left(\Delta F/F_{\rm em}\right)_1$ accurate to $O(v/c)$. 

Because of heavy dust obscuration, the S-stars are observed in the infrared. At these wavelength the emission of hot S-stars lies in the Rayleigh-Jeans tail of the blackbody spectrum, and we can safely assume $\alpha\approx 2$.

To compute flux variation as a function of time $t$ one needs to know how the stellar velocity $\mathbf{v}$ and distance $r$ relative to the BH vary with $t$. We assume ${\bf v}(t)$ and $r(t)$ to be given by their Keplerian values, neglecting the relativistic corrections (which provide a relative contribution at $O(v^2/c^2)$ order, negligible at our level of accuracy), e.g. the GR pericenter precession\footnote{The largest pericenter shift for the stars considered in this study does not exceed $2^\circ$ per orbit \citep{Peis2020b}.}.

As in \citet{Do2019}, we account for the Roemer delay --- the propagation time across the binary orbit (neglecting the Shapiro delay). The corresponding time correction is rather small (typically several days) as it is $\sim \langle v\rangle/c$ of the orbital period $P_{\rm orb}$, where $\langle v\rangle$ is the {\it mean} stellar velocity (which is considerably lower than the pericenter velocity for highly eccentric orbits). This time shift does not affect the {\it amplitude} of the Doppler boosting signal, which is of most interest for us.

%%%%%%%%%%%%%%%%%%%%%%%%%%%%%%%%%%
\begin{figure}
\centering
\includegraphics[width=0.48\textwidth]{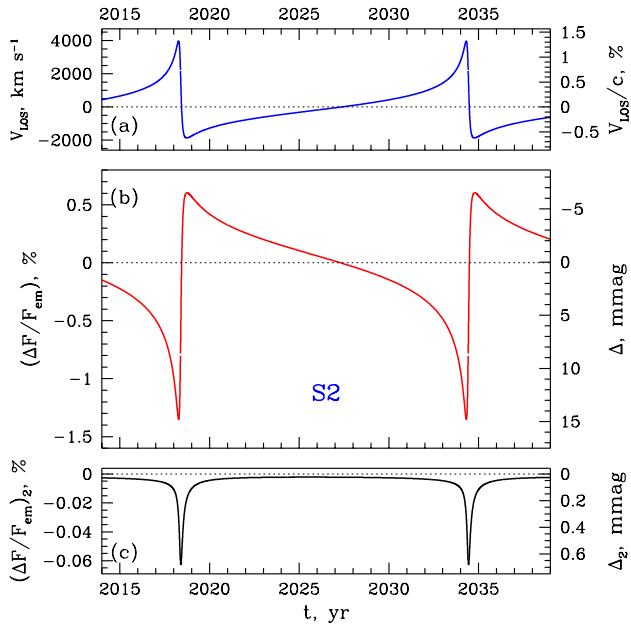}
\caption{
Velocity and photometric variability curves as a function of time for the S-star S2. We show the line-of-sight velocity $v_{\rm LOS}$ (a), full relative photometric variability due to the Doppler boosting $\Delta F/F_{\rm em}$ (b), and the contribution $\left(\Delta F/F_{\rm em}\right)_2$ due to the $O(v^2/c^2)$ effects --- gravitational redshift and transverse Doppler effect (c). Right axes in panels (b) and (c) measure flux deviations in mmag, while in panel (a) it shows $v_{\rm LOS}$ in units of $c$. 
\label{fig:S2}}
\end{figure}
%%%%%%%%%%%%%%%%%%%%%%%%%%%%%%%%%%

%%%%%%%%%%%%%%%%%%%%%%%%%%%%%%%%%%
\begin{figure}
\centering
\includegraphics[width=0.48\textwidth]{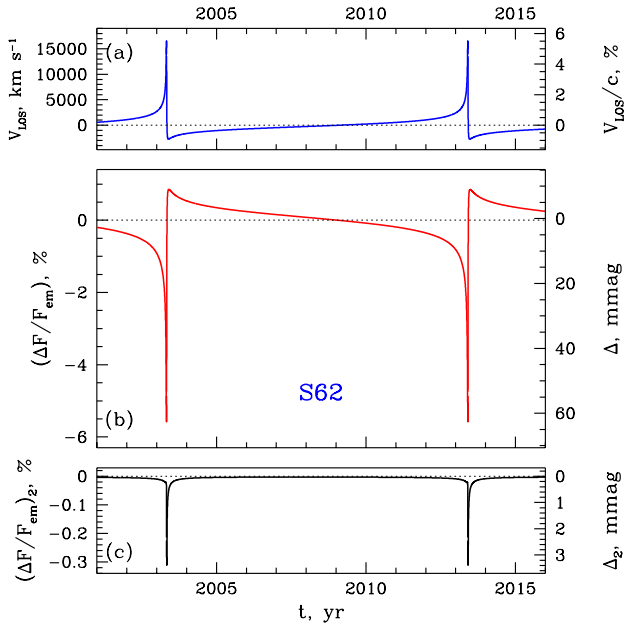}
\caption{
Same as Figure \ref{fig:S2} but for the S-star S62. 
\label{fig:S62}}
\end{figure}
%%%%%%%%%%%%%%%%%%%%%%%%%%%%%%%%%%

%%%%%%%%%%%%%%%%%%%%%%%%%%%%%%%%%%%%%%%%%%%%%%%%%%%%%%%%%%%
%%%%%%%%%%%%%%%%%%%%%%%%%%%%%%%%%%%%%%%%%%%%%%%%%%%%%%%%%%%

\section{Results}  
\label{sect:results}

%%%%%%%%%%%%%%%%%%%%%%%%%%%%%%%%%%%%%%%%%%%%%%%%%%%%%%%%%%%

\begin{figure}
\centering
\includegraphics[width=0.48\textwidth]{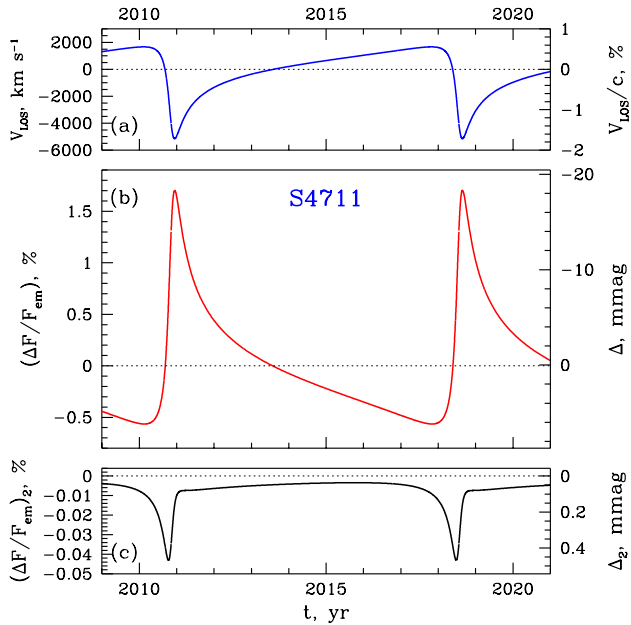}
\caption{
Same as Figure \ref{fig:S2} but for the S-star S4711. 
\label{fig:S4711}}
\end{figure}

\begin{figure}
\centering
\includegraphics[width=0.48\textwidth]{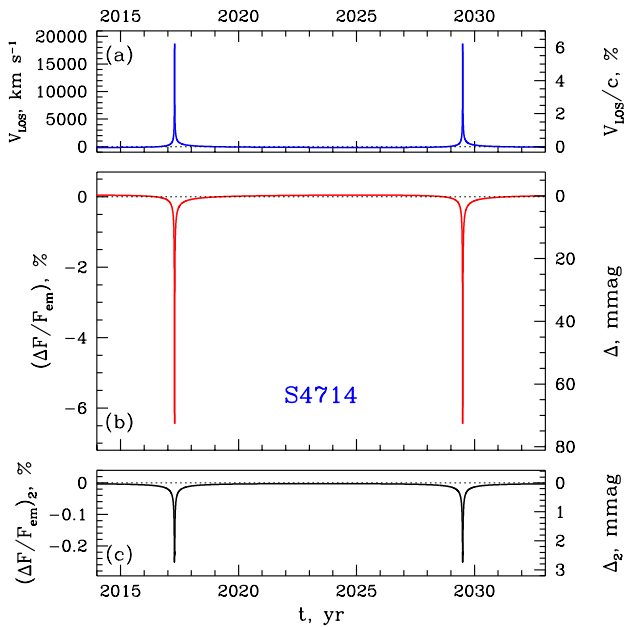}
\caption{
Same as Figure \ref{fig:S2} but for the S-star S4714. 
\label{fig:S4714}}
\end{figure}

We now describe the results of our calculation of the Doppler boosting signal for several S-stars in the Galactic Center, some of which have been recently discovered by \citet{Peis2020a,Peis2020b}. We choose four objects\footnote{Note that \citet{Gravity2020-S62} challenged the determination of the orbital parameters of S62 by \citet{Peis2020a,Peis2020b}, which may be considered as tentative.} most promising from the point of view of revealing observable photometric variation, including S2, with the properties listed in Table \ref{table:stars}. We illustrate their lightcurves (relative flux deviations in a narrow filter) computed using equation (\ref{eq:final}) in Figures \ref{fig:S2}-\ref{fig:S4714} (middle panels), in which we also plot their line-of-sight velocity $v_{\rm LOS}(t)$ (top panels). 

In the bottom panels of these figures we plot $(\Delta F/F_{\rm em})_2$, the photometric deviation due to $O(v^2/c^2)$ effects ---  gravitational redshift and transverse Doppler shift ---  defined as $(\Delta F/F_{\rm em})_2=(\Delta F/F_{\rm em})-(\Delta F/F_{\rm em})_1$. Using equations (\ref{eq:final}) and (\ref{eq:final1}) one can show for $\alpha=2$ that, to leading order, 
\ba   
(\Delta F/F_{\rm em})_2\approx \frac{1}{c^2}\left[\frac{v^2}{2}-\frac{GM_{\rm BH}}{r}-v_\perp^2\right],
\label{eq:final2}
\ea 
where $v_\perp$ is the transverse velocity (in the sky plane). The first two terms in the right hand side represent specific total energy of the star, which is negative for bound objects, implying that $(\Delta F/F_{\rm em})_2<0$. Note that $(\Delta F/F_{\rm em})_2$ represents the $O(v/c)$ {\it relative} correction to the leading-order result (\ref{eq:final1}), which is an $O(v/c)$ effects itself.

We now discuss our results for each individual star listed in Table \ref{table:stars} (its last two columns also list the ranges of variation of $\Delta F/F_{\rm em}$ and $(\Delta F/F_{\rm em})_2$).\\

\noindent {\bf S2}~~~
S2 is a well studied object with accurately determined orbital  parameters and a number of post-Newtonian effects --- GR redshift and transverse Doppler shift \citep{Do2019}, GR apsidal precession \citep{Abuter2020} --- measured using spectroscopy and astrometry. As shown in Figure \ref{fig:S2}, its $v_{\rm LOS}$ changes by almost 6000 km s$^{-1}$ near the pericenter (at $r_p=115$ au), which is reflected in the photometric signal of S2 changing by about $2\%$ over the time span of about half a year. The  $O(v^2/c^2)$ effects contribute to photometry at the level of $0.03\%$, making them difficult to observe.\\

\noindent {\bf S62}~~~
The discovery of S62 has been announced in \citet{Peis2020a}. Its orbital parameters (and $K$-magnitude of 16.1), as stated in \citet{Peis2020b}, make it quite favorable for revealing the photometric signature of the Doppler boosting. Near pericenter (at $r_p=17.8$ au, albeit with large uncertainty of $\pm 7.4$ au, \citealt{Peis2020b}) its line-of-sight velocity changes by over $19,000$ km s$^{-1}$ in about 27 d, causing its IR flux to change by about $6.4\%$, see Figure \ref{fig:S62}. The contribution to S62 photometric signal due to the GR redshift and transverse Doppler shift varies by $\approx 0.31\%$ during that time interval. \\

{\bf S4711}~~~
This object has the shortest orbital period around the Galactic Center BH as claimed by \citet{Peis2020b} --- about 7.7 yr, see Table \ref{table:stars}. However, the relatively low eccentricity places its pericenter at $r_p=142$ au, far enough from the BH to make $v_{\rm LOS}$ to change by only about $6800$ km s$^{-1}$ in the course of about 10 months. Over that period the photometric signal changes by $\approx 2.3\%$, while the $O(v^2/c^2)$ effects contribute to this variation at the level of $0.04\%$.\\

{\bf S4714}~~~ This star has the third shortest orbital period among the currently claimed S-stars (see Table \ref{table:stars}), but its very high eccentricity brings its pericenter to $r_p\approx 12.6$ au (although with large uncertainty, $\pm 9.3$ au, \citealt{Peis2020b}). This results in large variation of the photometric signal at the level of $6.5\%$, of which $0.25\%$ comes from $O(v^2/c^2)$ effects.

%%%%%%%%%%%%%%%%%%%%%%%%%%%%%%%%%%
%%%%%%%%%%%%%%%%%%%%%%%%%%%%%%%%%%

%%%%%%%%%%%%%%%%%%%%%%%%%%%%%%%%%%
%%%%%%%%%%%%%%%%%%%%%%%%%%%%%%%%%%

The orbit of S4714 has a rather peculiar orientation, with its apsidal axis lying almost in the plane of the sky, which explains the spiky behavior of $v_{\rm LOS}(t)$ and $\Delta F/F_{\rm em}$ in Figure \ref{fig:S4714}. For that reason, to characterize the timescale of photometric variability we do not use the time between the $v_{\rm LOS}$ extrema, which is about half of $P_{\rm orb}$. Instead, we use the time it takes $|\Delta F/F_{\rm em}|$ of S4714 to stay above its half-maximum value, which is very short, about 5.3 d. 

Unfortunately, due to its faintness (m$_K=17.7$, \citealt{Peis2020b}) this star may present less attractive target for the photometric followup than S62. 

%%%%%%%%%%%%%%%%%%%%%%%%%%%%%%%%%%%%%%%%%%%%%%%%%%%%%%%%%%%
%%%%%%%%%%%%%%%%%%%%%%%%%%%%%%%%%%%%%%%%%%%%%%%%%%%%%%%%%%%

\section{Discussion}  
\label{sect:disc}

%%%%%%%%%%%%%%%%%%%%%%%%%%%%%%%%%%%%%%%%%%%%%%%%%%%%%%%%%%%

Calculations presented in Section \ref{sect:results} clearly demonstrate that Doppler boosting can potentially provide an interesting purely photometric probe of the orbital dynamics in the Galactic Center. Another source of photometric variability of the S-stars could be the changing amplitude and orientation of their tidal bulge induced by the black hole gravity. The amplitude of this effect at the pericenter distance $r_p$ can be estimated as \citep{Morris1985}
\ba  
&& \left(\frac{\Delta F}{F}\right)_{\rm ell}\sim \frac{M_{\rm BH}}{M_\star}\left(\frac{R_\star}{r_{\rm p}}\right)^3
\label{eq:ell}\\
&&\approx 4\times 10^{-4} \left(\frac{M_\star}{M_\odot}\right)^{-1}
\left(\frac{R_\star}{R_\odot}\right)^{3}
\left(\frac{r_{\rm p}}{10~\rm{au}}\right)^{-3},
\nonumber
\ea  
where $M_\star$ and $R_\star$ are the stellar mass and radius. One can see that this effect is generally negligible compared to the Doppler boosting (its signal also drops faster with the distance from the BH).

%%%%%%%%%%%%%%%%%%%%%%%%%%%%%%%%%%%%%%%%%%%%%%%%%%%%%%%%%%%

\subsection{Applications}  
\label{sect:applications}

%%%%%%%%%%%%%%%%%%%%%%%%%%%%%%%%%%%%%%%%%%%%%%%%%%%%%%%%%%%

Measurement of the Doppler boosting signal can be used to confirm and refine orbital parameters of the faint S-stars closely approaching the black hole, for which astrometry and spectroscopy are not very accurate (if at all possible; GRAVITY interferometer on VLT may be particularly helpful for detecting new objects of this class, see \citealt{Gravity2020-S62}). It may be especially useful for better constraining the pericenter distance or eccentricity of the high-$e$ stars. According to \citet{Peis2020b}, the two stars with the smallest pericenter distances, S62 and S4714, have large error bars on their $r_p$ measurement, about 40\% and 70\%, correspondingly. If we change their eccentricities from the values listed in Table \ref{table:stars} in such a way that their $r_p$ would decrease by those amounts (i.e. to $e=0.986$ and $0.996$, correspondingly), then their variations of $\Delta F/F_{\rm em}$ would increase to $8.1\%$ for S62 and $11.9\%$ for S4714. The differences from the values reported in Figures \ref{fig:S62} and \ref{fig:S4714} are significant enough for the Doppler boosting alone to be able to place useful constraints on the pericenter distances. 

Dopper boosting must also be explicitly taken account when the photometry of S-stars is used for probing the circum-BH environment. \citet{Elaheh2020} found the intrinsic flux variability of S2 in the $L^\prime$ band to be $\approx 2.5\%$, allowing them to draw interesting conclusions about the circum-BH gas density structure. We find the photometry of S2 to vary due to Doppler boosting by $\approx 2\%$ in the IR (see Figure \ref{fig:S2}), strongly suggesting that this signal should be properly accounted for in future photometric studies of S2. For some other S-stars, e.g. S62 and S4714, accounting for the Doppler boosting is even more crucial\footnote{\citet{Elaheh2020} have already suggested using S62 as a probe of the gaseous environment of Sgr A$^*$.}, as it amounts to photometric variations larger than $6\%$, see Figures \ref{fig:S62} and \ref{fig:S4714}.

Our results also show that, if the photometric monitoring of the S-stars with relative accuracy of $10^{-4}-10^{-3}$ ever becomes possible, it may deliver a {\it purely photometric} measurement of the $O(v^2/c^2)$ effects --- gravitational redshift and transverse Doppler shift (see the bottom panels of our plots). It should be remembered, however, that $(\Delta F/F_{\rm em})_2$ signal could be comparable to the ellipsoidal variations, which would need to be explicitly modeled as well, see equation (\ref{eq:ell}).

%%%%%%%%%%%%%%%%%%%%%%%%%%%%%%%%%%%%%%%%%%%%%%%%%%%%%%%%%%%

\subsection{Observability}  
\label{sect:obs}

%%%%%%%%%%%%%%%%%%%%%%%%%%%%%%%%%%%%%%%%%%%%%%%%%%%%%%%%%%%

Photometric accuracy at the $\sim 1\%$ level necessary for detecting $\Delta F/F_{\rm em}$ computed in Section \ref{sect:results} should be achievable by current and future technology (\citealt{Elaheh2020} expect the future METIS spectrograph on the ELT to deliver 1 mmag precision for S2), although the faintness of the S-stars may require long integration times to beat down the photon noise. However, measuring Doppler boosting for the S-stars is likely to be a non-trivial exercise for reasons specific for the Galactic Center environment. 

First, the region around the Sgr A$^*$ is a very crowded field with stellar PSFs often overlapping each other. This could make assigning the observed flux to a particular star difficult, thus negatively impacting photometric accuracy. Careful modeling of the PSFs of multiple stars in the field (as well as the good understanding of the PSF itself, for which JWST may be needed) will likely be necessary to mitigate this issue.

Second, Sgr A$^*$ associated with the central BH is an infrared source itself. It is variable on a variety of timescales with its brightness typically fluctuating in the range 16-18 mag in K band \citep{Do2019Flare,Gravity2020flux}, which is comparable to the apparent magnitudes of many of the fainter S-stars, see Table \ref{table:stars}. The situation is complicated by the occasional flares, during which the Sgr A$^*$ flux can increase by more than an order of magnitude. This variability of the Sgr A$^*$ may present a serious issue for measuring the Doppler boosting signal as the stars with the largest $\Delta F/F_{\rm em}$ are also the ones that closely approach Sgr A$^*$, making it difficult to separate their photometric signals. Such separation ($r_{\rm p}=10$ au near the Sgr A$^*$ translates into $\sim 1$ mas at Earth) may be potentially achieved using the GRAVITY instrument on the VLT, which can reach $\lesssim 0.1$ mas resolution \citep{Gravity2017}. Also, if the IR flux variations of Sgr A$^*$ are mirrored in some other band (e.g. at longer wavelengths where the S-stars are dim, see  \citealt{Iwata2020}), then the simultaneous multi-frequency observations may allow one to subtract the Sgr A$^*$ contribution from the IR flux revealing the stellar Doppler boosting signal. 

Third, the emitted stellar flux $F_{\rm em}$ can also vary. However, its intrinsic variability due to rotation and star spots is likely not to be a serious issue, as stars hotter than $\approx 8000$ K tend to show low levels of flux variation, often $\lesssim 1$ mmag  \citep{Balona2013,Balona2017,Savanov2019}, which is well below $\Delta F/F_{\rm em}$ found in this work (unless one tries to also measure $O(v^2/c^2)$ effects). 

Finally, spatially and temporally variable dust extinction in the Galactic Center region can introduce spurious trends in the photometry of S-stars on long time scales. However, it should be possible to mitigate this issue by measuring reddening in multiple bands.

%%%%%%%%%%%%%%%%%%%%%%%%%%%%%%%%%%%%%%%%%%%%%%%%%%%%%%%%%%%
%%%%%%%%%%%%%%%%%%%%%%%%%%%%%%%%%%%%%%%%%%%%%%%%%%%%%%%%%%%

\section{Summary}  
\label{sect:summary}

%%%%%%%%%%%%%%%%%%%%%%%%%%%%%%%%%%%%%%%%%%%%%%%%%%%%%%%%%%%

We explored the importance of Doopler boosting for the photometric observations of S-stars in the Galactic Center. Focusing on four objects with very tight orbits around the BH, we demonstrated that Doppler boosting leads to photometric variations of these S-stars at the level of several per cent (around $6\%$ for S62 and S4714) around the time of pericenter passage. Measurement of this signal can be used for confirming and refining orbital parameters of the S-stars for which spectroscopy and astrometry are difficult. Doppler boosting should also be accounted for when S-stars are used as probes of the gaseous environment of the Sgr A$^*$. Intrinsic variability of Sgr A$^*$ and crowding of the surrounding stellar field are the biggest observational challenges that a successful measurement of Doppler boosting would need to overcome. Measuring higher order, $O(v^2/c^2)$, effects due to transverse Doppler shift and GR redshift is unlikely to be be possible unless the systematic effects discussed in Section \ref{sect:obs} are suppressed by orders of magnitude. 

\acknowledgements

I am grateful to Zoltan Haiman for useful comments and to Eric Ford for advice on stellar variability. Financial support for this work was provided by the NASA grant 15-XRP15-2-0139, STFC grant ST/T00049X/1, and John N. Bahcall Fellowship. 

%%%%%%%%%%%%%%%%%%%%%%%%%%%%%%%%%%%%%%%%%%%%%%%%%%%%%%%%%%%
%%%%%%%%%%%%%%%%%%%%%%%%%%%%%%%%%%%%%%%%%%%%%%%%%%%%%%%%%%%

%%%%%%%%%%%%%%%%%%%%%%%%%%%%%%%%%%%%%%%%%%%%%%%%%%%%%%%%%%%
%%%%%%%%%%%%%%%%%%%%%%%%%%%%%%%%%%%%%%%%%%%%%%%%%%%%%%%%%%%

\bibliographystyle{apj}
\bibliography{references}

\end{document}